\documentclass[aps,prl,preprint,showpacs,amsmath,amssymb]{revtex4}

\ifx\pdftexversion\undefined
  \usepackage[dvips]{graphicx}
\else
  \usepackage[pdftex]{graphicx}
\fi

\usepackage{dcolumn}
\usepackage{bm}

\begin{document}

\title{Reduced electron relaxation rate in multi-electron quantum dots}

\author{Andrea Bertoni}
\affiliation{INFM National Research Center on nanoStructures and
bioSystems at Surfaces (S3)}
\affiliation{Dipartimento di Fisica,
Universit\`a degli Studi di Modena e Reggio Emilia, Via Campi 213/A, 41100
Modena, Italy}

\author{Massimo Rontani}
\affiliation{INFM National Research Center on nanoStructures and
bioSystems at Surfaces (S3)}
\affiliation{Dipartimento di Fisica,
Universit\`a degli Studi di Modena e Reggio Emilia, Via Campi 213/A, 41100
Modena, Italy}

\author{Guido Goldoni}
\affiliation{INFM National Research Center on nanoStructures and
bioSystems at Surfaces (S3)}
\affiliation{Dipartimento di Fisica,
Universit\`a degli Studi di Modena e Reggio Emilia, Via Campi 213/A, 41100
Modena, Italy}

\author{Elisa Molinari}
\affiliation{INFM National Research Center on nanoStructures
and bioSystems at Surfaces (S3)}
\affiliation{Dipartimento di Fisica,
Universit\`a degli Studi di Modena e Reggio Emilia, Via Campi 213/A, 41100
Modena, Italy}

\date{\today}

\begin{abstract}
We use a configuration-interaction approach  and Fermi golden rule  to
investigate electron-phonon   interaction in  realistic multi-electron
quantum  dots. Lifetimes   are computed  in  the low-density,   highly
correlated regime. We report numerical evidence that electron-electron
interaction  generally leads   to  reduced  decay  rates of    excited
electronic states  in  weakly  confined  quantum  dots, where  carrier
relaxation is dominated by  the interaction with longitudinal acoustic
phonons.
\end{abstract}


\maketitle

Semiconductor quantum  dots  (QDs)  are artificially-confined  systems
where full control on the number of  carriers and energy states can be
achieved by either tuning the confinement  potential or using external
fields   \cite{jacakbook}.  These    confined   systems    can   reach
unprecedented regimes of large  Coulomb-to-kinetic energy ratios, thus
offering an ideal   laboratory   for the exploration of    correlation
effects at the nanoscale  \cite{reimann02}. In addition, QD devices in
the single-electron (SE) or multi-electron  (ME) regime hold a primary
role  in new classes of  applications, ranging from optoelectronics to
biology and quantum information processing \cite{librofaustoSQUID}.

One   fundamental question is how the   discrete density of electronic
states affects  the   relaxation  time  of   excited  electrons.   The
knowledge of  carrier lifetime  is  obviously of great importance  for
many  applications,  and most  significant  for  the implementation of
solid-state gates for quantum information  processing, which require a
large number of elementary operations to  be realized within the phase
coherence time   of the relevant degrees  of  freedom.   It  was early
recognized that, since scattering from optical phonon is suppressed in
weakly bound QDs (with confinement energy below $\sim 10$ meV in GaAs)
\cite{inoshita92,notaSTRONGCONF}, the coupling  with  the
longitudinal acoustic (LA) phonon continuum becomes the main source of
scattering, thus    limiting  charge  relaxation times    to   tens of
nanoseconds  \cite{bockelmann94}.  Theoretical  estimates of LA-phonon
induced scattering rates  were obtained in a SE  approach and showed a
satisfactory agreement with transport-spectroscopy experiments
\cite{fujisawa02}.   However, the  technologically relevant problem of
estimating  the carrier  lifetime  when more   than  one electron  are
present in the device   and, in general,  the fundamental  question of
whether Coulomb interaction between carriers affects the relaxation of
ME quantum  states have received little  attention~\cite{brasken02}.

In  this  Letter  we report  numerical   evidence that carrier-carrier
Coulomb interaction leads to  reduced  relaxation rates of excited  ME
states in weakly confined, strongly correlated QDs when the dominating
intra-band   scattering   mechanism    is    the   interaction    with
LA-phonons. The lifetime for a specific QD is shown to strongly depend
on the number of carriers. 

The calculation of  ME states in  QDs  has been approached by  several
methods (for a review of quantum states in single  and coupled QDs see
Refs.~\onlinecite{reimann02}  and  \onlinecite{rontani01}).  Since  we
are interested in the relaxation time of excited ME states, we need to
know  both ground and  excited states  with  comparable accuracy.  Our
method of choice is  the configuration interaction (CI) approach:  the
correlated wave functions of the ground and excited states are written
as       superpositions        of     Slater    determinants    (SDs),
$\left|\Phi_i\right>=\prod_{\alpha_i}\!c^\dag_{\alpha_i}\!
\left|0\right>$, obtained by  filling in the SE spin-orbitals $\alpha$
with  the       $N$  electrons      in \emph{all}      possible   ways
($c^{\dagger}_{\alpha}$ creates an electron   in level $\alpha$).  The
fully interacting  Hamiltonian,  written within the  envelope function
and effective mass    approximations,  is first  block   diagonalized,
exploiting orbital and  spin symmetries.  Finally, the  ME Hamiltonian
is diagonalized   in  each symmetry  sector, giving   the full set  of
excited states of the system~\cite{rontani04}.  

In order to  evaluate   quantitatively the  relaxation time   at  zero
temperature between given ME states  $b$ and $a$,  $\tau_{b\rightarrow
a}$, we  employ the    Fermi  golden rule between    fully  correlated
states~\cite{notaooe},
\begin{equation} \label{eq1}
\tau^{-1}_{b\rightarrow a}
= \frac{2\pi}{\hbar} \sum_{\bm{q}} \left|
\sum_{i j} c^*_{bi} c_{aj} \langle\Phi_i|V_{\bm{q}}|\Phi_j\rangle
\right|^2 \delta(E_b-E_a-\hbar\omega_{\bm{q}}),
\end{equation}
where the ME states $|\Psi_a\rangle=\sum_i c_{ai} |\Phi_i\rangle$ have
been   written   explicitly    as    linear    combinations  of    SDs
$|\Phi_i\rangle$,   $V_{\bm{q}}$ is the   interaction  operator of  an
electron with  a LA-phonon  with   wave vector  ${\bm{q}}$,  $E_a$ and
$\hbar\omega_{\bm{q}}$  are    the   ME-state   and  phonon  energies,
respectively. We use   the standard  deformation-potential  model  for
$V_{\bm{q}}$     and  a linear  dispersion   for   the  bulk LA-phonon
branch~\cite{ferrybook91,noterella}.

The  SE levels are calculated for  a  cylindrical single or double QD,
with    confinement       potential     $V({\bm{r}})=      V_z(z)    +
\frac{1}{2}m^*\omega_0^2(x^2+y^2)$, $V_z(z)$ representing the vertical
confinement of a single or  double quantum well, $\hbar\omega_0$ being
the   SE energy level spacing  of   a lateral two-dimensional harmonic
trap,   and  $m^*$ being the   effective  mass.   We  consider typical
``vertical''   GaAs/AlGaAs    samples    with   effective  cylindrical
geometry~\cite{tarucha96,ashoori96}.     For this  configuration,  and
neglecting spin-orbit coupling, the  SE spin-orbitals can be given the
separable form $\psi_{nmg\sigma}({\bm{r}},s)= \varphi_{nm}(x,y)
\phi_{g}(z)\chi_{\sigma}(s)$,   with $s=\uparrow,\downarrow$ the spin
state,  $n=0,1,\dots$ the   radial and  $m=0,\pm 1,\pm   2, \dots$ the
angular   quantum  numbers  related to  the   lateral confinement, and
$g=0,1,\dots$ labeling the eigenstate in the $z$ direction.  Since the
vertical confinement is much stronger   than the  lateral one, it   is
sufficient to consider    $g=0$ for single QDs,  and   $g=0,1$ for QDs
coupled  via vertical  tunneling.    In  the  ME Hilbert  space,   the
resulting global symmetries are   the total orbital  angular  momentum
$z$-component $M$,  total     parity $G$, total  spin   $S$,   and its
$z$-component  $S_z$,  which are   used  to block  diagonalize  the ME
Hamiltonian. This  numerical scheme has   proved extremely reliable in
reproducing the measured   transport and  inelastic  light  scattering
spectra in single and coupled QDs~\cite{rontani04,garcia05}.

Equation (\ref{eq1}) shows   explicitly that  the ME scattering   rate
$\tau^{-1}_{b\rightarrow a}$ can be thought of  as a combination of SE
scattering   events, where a given configuration   (i.e., a single SD)
changes  at  most   by  the   occupation   of   a single  spin-orbital
$\psi_{nmg\sigma}(\bm{r},s)$.    The  coefficients    $c_{ai}$     are
determined by  the solution of  the interacting problem through the CI
method.   It   should   be  noted   that   the SE      matrix elements
$\langle\Phi_i|V_{\bm{q}}|\Phi_j\rangle$ depend strongly   \emph{both}
on the SE orbitals $\varphi_{nm}(x,y)\phi_g(z)$, possibly changing via
confinement potential  and  external fields, \emph{and} on  the phonon
wave vector $\bm{q}$.  The magnitude  of $\bm{q}$ is determined by the
transition energy $E_b-E_a$, fixed, in turn, by the difference between
the SE energies of  initial and final  levels plus the Coulomb energy.
Indeed, typical  LA-phonon wavelengths are  comparable to  the spatial
extension of SE quantum  states.  Coulomb interaction, therefore,  may
affect  $\tau^{-1}_{b\rightarrow a}$ through  both the CI coefficients
$c_{ai}$  and  the SE  matrix  elements  via shift  of  the transition
energies $E_b-E_a$.

\begin{figure}[t]
\includegraphics[clip=true,scale=.28]{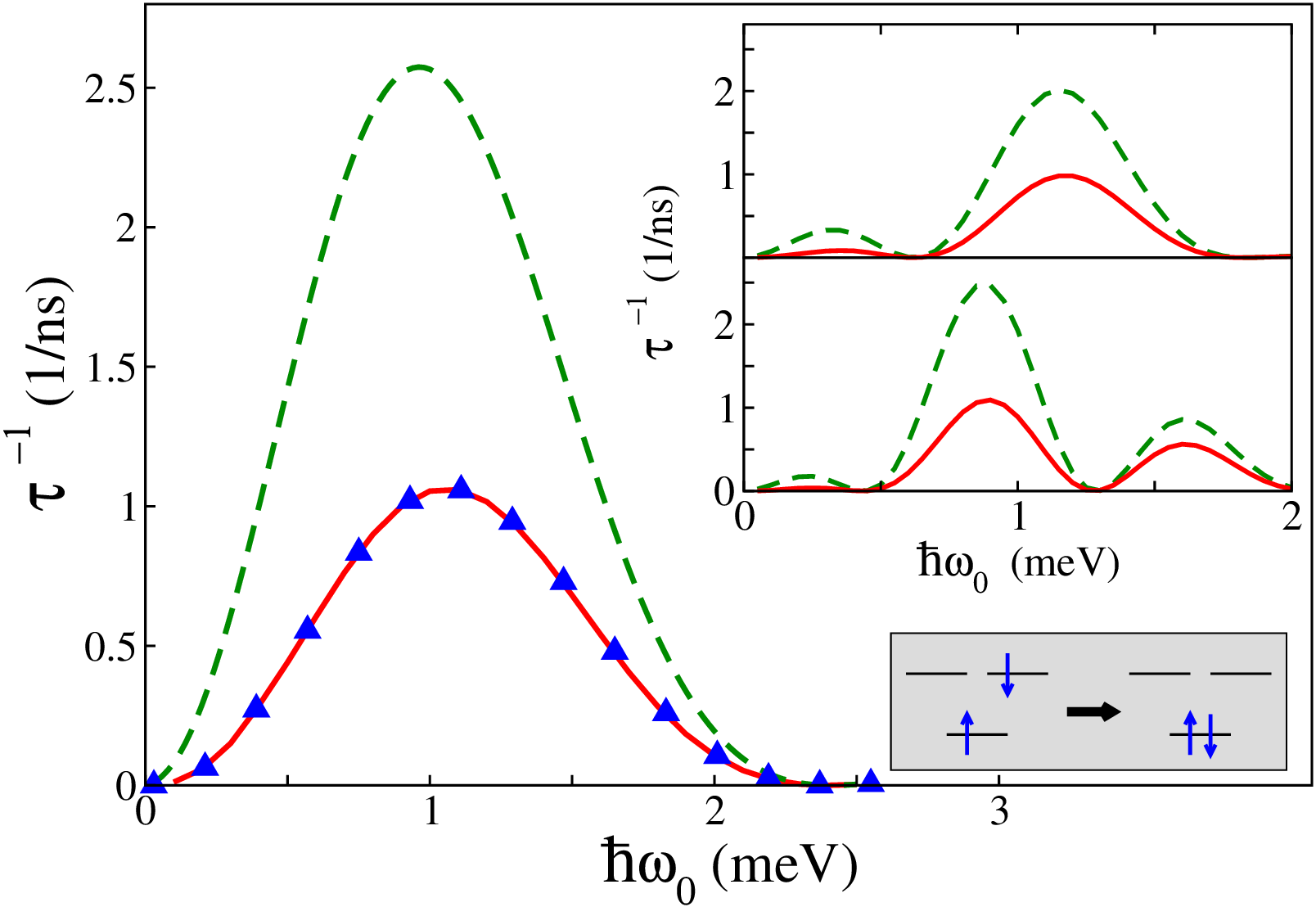}
\caption{\label{fig1}  Electron-LA phonon scattering rate at T=0 for:
\emph{dashed curve}, SE $p\rightarrow s$ transition, and,
\emph{solid curve}, two-electron  $(M=1,S=0)\rightarrow (M=0,S=0)$
 transition  (see  lower inset),   for a cylindrical  QD   (GaAs layer
 $10$~nm  thick), as a   function  of the lateral  confinement energy,
 $\hbar\omega_0$.  Scattering rates for one and two electrons would be
 the  same  if correlation  were   switched off.   Triangles show  the
 two-electron rate calculated by artificially taking the energy of the
 emitted phonon  equal to the confinement  energy, as in  the SE case:
 the values  do not differ  significantly from the ``true'' transition
 (solid line).  The  upper  insets show the   corresponding transition
 rates  for a double dot  with inter-dot  barrier $3$~nm (upper graph)
 and $8$~nm (lower graph) wide, respectively.  }
\end{figure}

We start  our discussion of ME-state  relaxation times  by considering
the two-electron case. In Fig.~\ref{fig1}  we show the relaxation rate
$\tau^{-1}_{b\rightarrow a}$  (solid curve) for the  LA-phonon induced
transition as a function  of the confinement energy $\hbar\omega_0$ in
a single QD, where $b\equiv(M=1,S=0)$ is  the correlated first excited
state   and $a\equiv(M=0,S=0)$ the ground   state  within the  singlet
sector.  This  transition is mostly composed  of a SE $p\rightarrow s$
transition  [$p\equiv(m=1,n=0)$,   $s\equiv(m=0,n=0)$],       with   a
``spectator'' electron in level $s$  (see diagram in Fig.~\ref{fig1}).
For comparison,     we show  in   Fig.~\ref{fig1}   the  corresponding
non-interacting   $p\rightarrow     s$     relaxation   rate   (dashed
curve)~\cite{bockelmann94}.  The relaxation   rate  of the  correlated
state is suppressed   by a factor  $\sim 2.5$  with respect  to the SE
transition.  The two panels  of the inset  in Fig.~\ref{fig1} plot the
same  transition  for the case of   a double-dot sample  with selected
values of the inter-dot barrier.  Here, the SE orbitals are coherently
delocalized across the sample,  with levels $(mng)=(100)$  and $(000)$
corresponding  to $p$ and $s$,  respectively.  The behavior is similar
to the single QD  system, with $\tau^{-1}_{b\rightarrow  a}$ vanishing
at the  low  and  high  energy   sides, but oscillations  appear    at
intermediate  values  of $\hbar\omega_0$;   these oscillations,  which
increase in number with the width of the inter-dot barrier, are a pure
SE effect and  are due to the  matching between the average separation
of   the         two          QDs     and           the         phonon
wavelength~\cite{bertoniAPL04,zanardi98,bertoniPHYSE05}.    We  stress
that, in the ME case, even  if $\tau^{-1}_{b\rightarrow a}$ is reduced
by  electron  correlation   with   respect to  the   corresponding  SE
transition, it  behaves qualitatively in  the same way.  Specifically,
the values of the  confinement energy and of  external fields at which
the scattering rate is suppressed (cf.~Fig.~\ref{fig1}) are the same.

Now  consider the lowest-energy  SE transition in   a parabolic QD, in
which case  the SE gaps   $\hbar\omega_0$ fix  the  energy of  emitted
(bulk)   phonons  $\hbar\omega_{\bm{q}}=\hbar\omega_0$.      As    the
confinement energy changes, e.g., by tuning the gate potential,
\emph{both}  the phonon energy   \emph{and} the SE orbitals vary.
In the SE case the scattering rate variation is due to the combination
of    these  two effects.   For   the    two-electron  case,  however,
$\hbar\omega_{\bm{q}}$  is in  general different from  $\hbar\omega_0$
due  to Coulomb contribution.   Therefore, in  order to verify whether
$\tau^{-1}_{b\rightarrow a}$   is reduced   by a  genuine  correlation
effect or  simply  by  the change   in the  emitted  phonon energy, we
calculated   $\tau^{-1}_{b\rightarrow  a}$   by   artificially  fixing
$\hbar\omega_{\bm{q}}$ at    the SE gap  $\hbar\omega_0$ also   in the
two-electron  case.  This is  done by substituting  the Dirac delta in
Eq.~(\ref{eq1}) with $\delta(\hbar\omega_0-\hbar\omega_{\bm{q}})$.  We
found  that the scattering rate   obtained in  this way (triangles  in
Fig.~\ref{fig1}) differs from the two-particle correlated case by only
a     few   percents.       Consequently,     the     reduction     in
$\tau^{-1}_{b\rightarrow  a}$  is brought  about  by correlation only,
since  the transition energy and,  therefore, the  emitted phonon wave
vector are only slightly affected by the interaction.

We show next that the  reduction of the  scattering rate brought about
by Coulomb correlation is a  more general effect and  holds also for a
larger  number of electrons.   In the  CI  approach, the ME transition
rate between two specific states  is the square   of a sum of many  SE
transition   amplitudes  [Eq.~(\ref{eq1})] and,  due   to interference
effects among those amplitudes, there  is no reason to believe \emph{a
priori} that the ME transition rates are larger or smaller than the SE
ones.  We found, however,  that the calculated  ME transition rate is,
in  general, smaller  than the  largest   transition rate between  two
uncorrelated states of the QD, as discussed below.

\begin{figure}[t]
\includegraphics[clip=true,scale=.26]{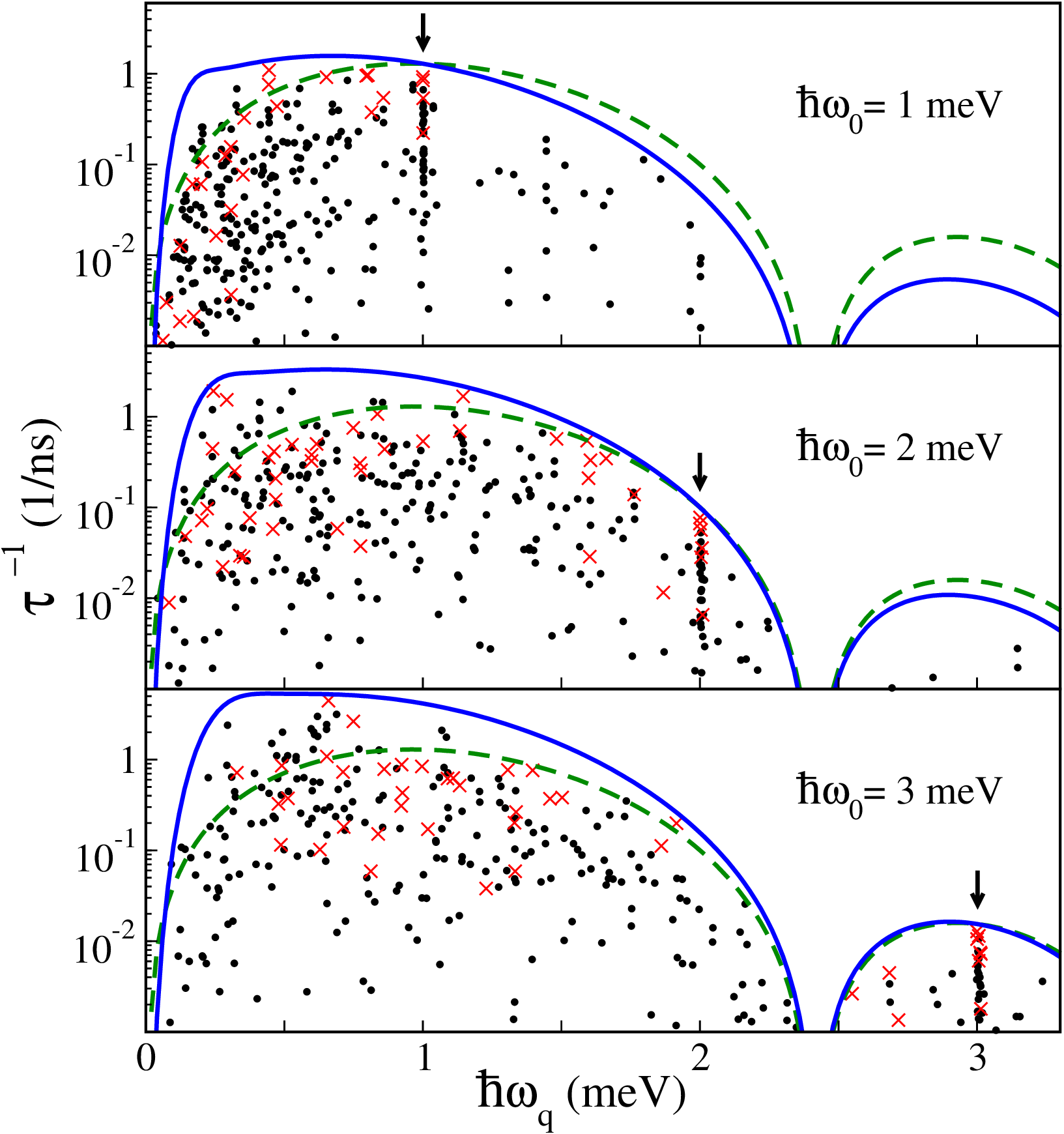}
\caption{\label{fig2}  Electron-LA phonon scattering rate, for three
QDs with  different  lateral confinement energy  $\hbar\omega_0$, as a
function of the energy  of the emitted phonon, $\hbar\omega_{\bm{q}}$.
Dashed curves  show  the   SE  scattering  rate  of the   lowest-energy
transition ($p\rightarrow  s$). 
Solid curves  refer  to  the same transition   except
orbitals are   fixed  to  the   value  of  $\hbar\omega_0$ indicated in
figure.  Dots  and
\textsf{X} symbols identify  specific  ME transitions from  excited to
ground states with $2\le  N \le 6$  (the first  20 excited states  for
each  $N$ were considered), with  the  \textsf{X} symbols labeling ME
transitions evolving into the  non-interacting ones as correlation  is
switched  off  (see text). Arrows identify Kohn modes.}
\end{figure}

In order to discuss the scattering rates for $N>2$, in Fig.~\ref{fig2}
we show the LA-phonon induced scattering rate $\tau^{-1}_{b\rightarrow
a}$ as a function  of the emitted phonon energy $\hbar\omega_{\bm{q}}$
in QDs at selected lateral confinements and $2\le N\le 6$.  Each point
in  the  graphs represents  a transition   between two  correlated  ME
states.  The first  20 eigenstates  are considered,  for each value of
$N$; due to   total  spin conservation,  only  a few  of  all possible
transitions   are   allowed.  In each     panel several ME transitions
accumulate   precisely at   $\hbar\omega_0$,   identified by  downward
arrows: these transitions   correspond to  the  de-excitation of  Kohn
modes  \cite{jacakbook},  i.e., the separable  center-of-mass modes of
the ME system, whose energies are exact multiples of $\hbar\omega_0$.

To  compare the  ME   scattering rates with   SE ones,   we  report in
Fig.~\ref{fig2} (dashed  curve) the  scattering   rate of the   lowest
($p\rightarrow s$) SE transition for a  QD whose parabolic confinement
energy matches  the emitted-phonon    energy,  as implied by  the   SE
picture.  Note that,  for such a curve,  the  graph abscissa indicates
also the confinement  energy.  Since  each ME  transition has  many SE
components, in  order   to  identify the  transitions   that are  more
directly  comparable with  the SE  ones,   we have indicated with  the
symbol \textsf{X} those  processes whose highest-weight components  of
initial  and final ME  states can  be  mapped one  into  the  other by
destroying an  electron occupying orbital $p$  (initial SD) and moving
it to  $s$ (final   SD): the  relaxation   time of these  specific  ME
transitions matches that of   the SE $p\rightarrow s$ relaxation  once
the correlation effects are neglected.

At the low-energy side of each  plot we see that  a few ME transitions
have a scattering rate which is higher than the SE one, which seems to
contradict results for $N=2$.  However,  note that here we are showing
the ME scattering rates  as a function  of $\hbar\omega_{\bm{q}}$ at a
given  $\hbar\omega_0$, while   in  the SE  case  (dashed   curve) the
confinement energy is $\hbar\omega_{\bm{q}}$, as  explained above.  To
single   out  the   specific  effects  of    Coulomb   interaction  on
$\tau^{-1}_{b\rightarrow a}$, a more significant comparison is with SE
transitions  (solid  curve)  calculated with   the lateral confinement
energy $\hbar\omega_0$ used for the  ME calculations (values indicated
in Fig.~\ref{fig2}), i.e.   releasing  the condition  that  equals the
harmonic confinement  to the phonon energy.  In other words  it is the
solid curve  of Fig.~\ref{fig2} that  corresponds to the $p\rightarrow
s$ SE transition contributing to the  $ij$ summation of Eq.~\ref{eq1}.
As      expected,   the  solid  and    dashed      curves  coincide at
$\hbar\omega_{\bm{q}}=\hbar\omega_0$.   For all considered cases,  the
scattering rate of ME transitions lies below the  SE one with the same
SE wave function and  phonon energy; this  is true, in particular, for
Kohn modes.

The  three QDs  considered   in Fig.~\ref{fig2} correspond  to  rather
different Coulomb-to-kinetic energy ratios \cite{egger99}: in terms of
$\lambda=l_0/a^*_{\text{B}}$,  where $l_0=\sqrt{\hbar/m^*\omega_0}$ is
the   QD  radius  and  $a^*_{\text{B}}$ the     effective Bohr radius,
$\lambda=3.3, 2.3,  1.9$  for  $\hbar\omega_0=1, 2, 3$   meV  and GaAs
parameters, respectively.  The  larger  $\lambda$ (and  the dot),  the
stronger  the    effect    of   correlation,  keeping     $N$    fixed
\cite{rontanimolly04}.  
Figure  \ref{fig2} shows  that   larger QDs with  stronger correlation
effects  have   lower   scattering   rates.   The   above   discussion
demonstrates that Coulomb interaction tends to decrease the scattering
rate of ME states  in weakly confined QDs.   From the point of view of
specific  applications,  if a low  SE  LA-phonon  scattering  rate  is
predicted for a specific sample  and/or field configuration, then such
desirable property will not be spoiled by Coulomb interactions.

\begin{figure}[t]
\includegraphics[clip=true,scale=.28]{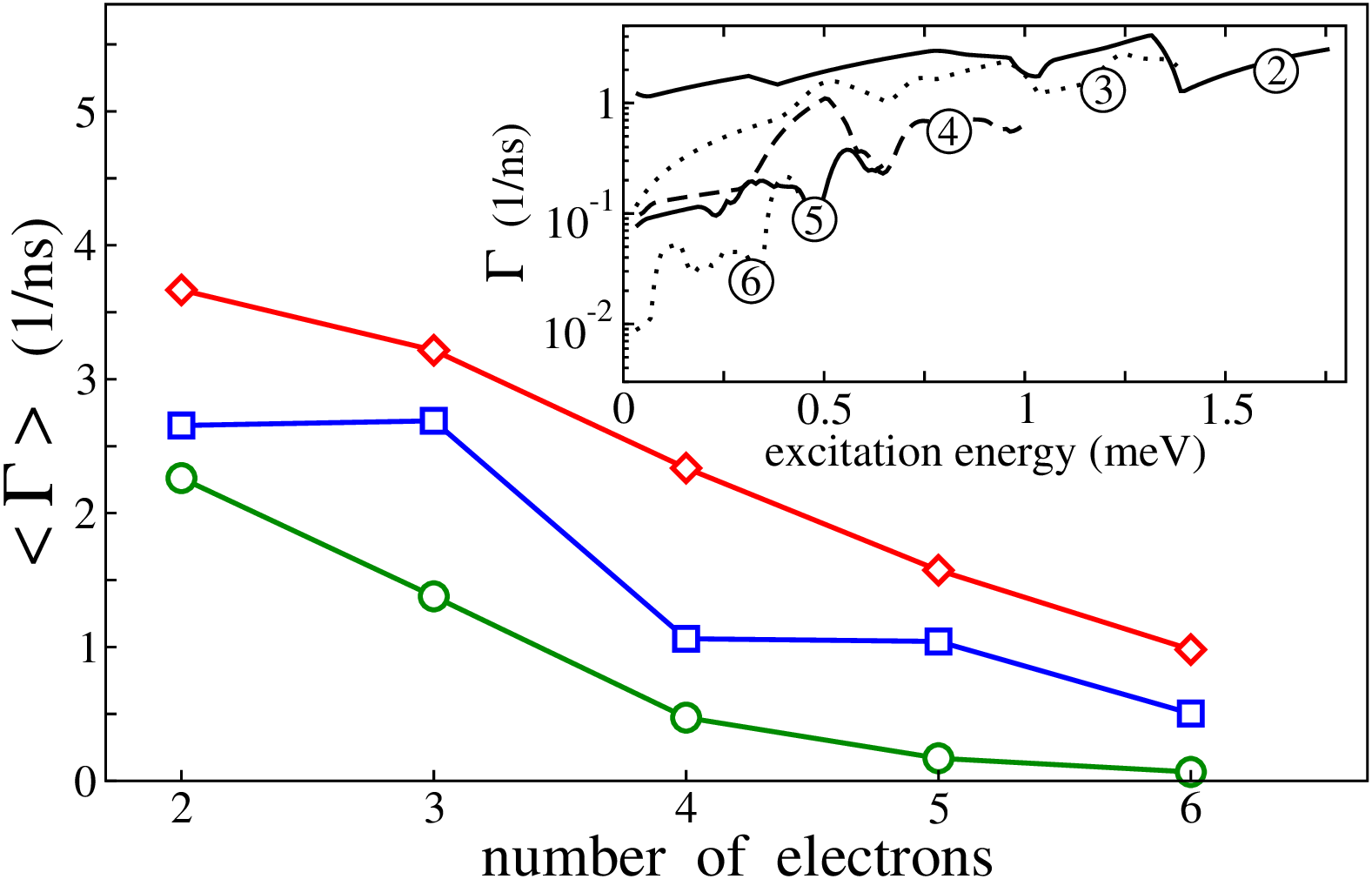}
\caption{\label{fig3}  Mean decay rate of correlated ME states with up
to six electrons, due to electron-LA  phonon scattering, for the three
QDs of Fig.~\ref{fig2}, with $\hbar\omega_0= 1$ meV (circles), $2$ meV
(squares), $3$  meV (diamonds).   Inset: decay  rate  of the first  20
correlated ME states as a function of the excitation energy for the QD
with $\hbar\omega_0= 1$  meV.   The curves connecting  the transitions
considered are guides to the eye.   The labels indicate the number of
electrons.
}
\end{figure}

The  lifetime $\Gamma^{-1}$ of a given  quantum state  depends both on
the   scattering rates  and  on the  number   of channels  of possible
de-excitation,   i.e.,  the  density  of   states  of symmetry-allowed
(spin-conserving)  final states.   In  the ME  system  under study, an
increasing number of  particles leads to a larger  number  of states a
given initial  state can decay  into.   Such effect  competes with the
above prediction     of  smaller scattering   rates.    We  report  in
Fig.~\ref{fig3} inset the inverse lifetime of the  first 20 QD excited
states for fixed $N$, considering  different curves for $2\le N\le 6$.
First we note  that, due to the varying  density of states, the energy
range covered by the transitions considered decreases as the number of
electrons increases. The 20 decay times are reported, for each $N$, as
a function of the difference between initial and final state energies,
where  the final state is the  $N$-electron ground  state.  The points
are connected with  lines as a guide to  the eye.  Although the curves
do not show a  monotonic behavior, two  global features emerge: (i) as
the energy of  the initial state increases,  its lifetime  lowers (ii)
the  lifetime  of a  quantum  state  increases  with $N$, provided the
excitation  energy is fixed.  Property (i)  holds since the density of
de-excitation channels increases with  the initial state energy, while
(ii) is due to the fact  that scattering rates decrease as correlation
effects become stronger. In the samples we considered, the effect (ii)
is not obscured  by (i).  This is shown  in Fig.~\ref{fig3}, where the
scattering time  averaged   over the  first  20  ME excited  states is
reported as a function of $N$ for three different QD samples.

To   summarize, we   quantitatively estimated  the   LA phonon-induced
relaxation  times of correlated ME  excited  states in weakly confined
cylindrical   GaAs/AlGaAs   QDs.   We   find  that   electron-electron
interaction contributes to  reduce the scattering rate. Furthermore we
showed  that  the  correlated   ME  energy   spectra  mimic   that  of
non-interacting QDs.   In  particular, the  transition  probability is
suppressed  for  phonon   energies  whose  wavelength matches  the  QD
vertical   dimension  (when  emission    is  inhibited  also  in   the
non-correlated  case   \cite{bertoniAPL04}),  independently  from  the
number of interacting electrons.  We believe that these effects are of
considerable relevance for the modeling of future nanodevices based on
coherent electron dynamics and, in  particular, for quantum  computing
proposals.

We  acknowledge support from the   Italian Ministry of Foreign Affairs
(DGPCC), the Italian Ministry of  Research under the FIRB (RBAU01ZEML)
and  COFIN (2003020984)  programs,   and INFM  Iniziativa  Trasversale
Calcolo Parallelo 2005. We are grateful to F. Troiani for discussions.

%

\newpage

\newpage

\newpage

\begin{thebibliography}{24}
\bibitem[{not({\natexlab{a}})}]{notaSTRONGCONF}
\bibinfo{note}{Due to the low lateral  confinement, the electron-phonon
  coupling is weak  in  the samples  considered.  As a  consequence  we
  neglect  the  second-order process in  which  the  emitted phonon is
  reabsorbed.  For  QDs in  the  strong coupling regime,  where a full
  diagonalization of the Fr\"olich Hamiltonian  is in order to  describe
  the polaron formation, see, for example,
  Refs. \onlinecite{hameau99,vagov03,hohenester04}.}
\bibitem[{not({\natexlab{b}})}]{notaooe}
\bibinfo{note}{Since we are considering an out-of-equilibrium process
at $T\sim0$, whose rate could be estimated by a transport pump-and-probe
experiment like the one reported in Ref.~\onlinecite{fujisawa02}, we
neglect the second-order ``in'' scatterings.}
\bibitem[{not({\natexlab{c}})}]{noterella}
\bibinfo{note}{Well converged results are obtained for the lowest 20 ME states
  for $1\le N \le 6$ using 56 SE spin-orbitals, corresponding up to $\approx
  3\times 10^5$ SDs.}
\end{thebibliography}
\end{document}